\documentclass[10pt,twocolumn,twoside]{IEEEtran}
\newcommand{\beq}{\begin{equation}}
\newcommand{\eeq}{\end{equation}}
\newcommand{\bsq}{\begin{subequations}}
\newcommand{\esq}{\end{subequations}}
\newcommand{\bq}{\begin{eqnarray}}
\newcommand{\eq}{\end{eqnarray}}
\newcommand{\bqn}{\begin{eqnarray*}}
\newcommand{\eqn}{\end{eqnarray*}}
\usepackage{cite}
\usepackage{url}
\usepackage{empheq}
\usepackage{float}
\usepackage{dsfont}
\usepackage[hidelinks]{hyperref}

\usepackage{threeparttable}
\usepackage{algorithm}
\usepackage{algorithmic}

\usepackage[table,xcdraw]{xcolor}

\usepackage{arydshln}
\usepackage{nicematrix}

\usepackage{amsmath,amssymb,amsfonts}
\allowdisplaybreaks[4]
\usepackage{graphicx}
\usepackage{textcomp}
\usepackage{amsmath}
\usepackage{enumerate}
\usepackage{epstopdf}
\usepackage{array}
\usepackage{booktabs}
\usepackage{subfigure}
\usepackage{multirow}
\usepackage{soul, color}
\usepackage[latin1]{inputenc}
\usepackage{cases}

\newtheorem{theorem}{Theorem}
\newtheorem{lemma}{Lemma}

\newtheorem{proposition}{Proposition}

\newtheorem{definition}{Definition}

\soulregister\cite7 
\soulregister\citep7 
\soulregister\citet7 
\soulregister\ref7 
\soulregister\pageref7

\usepackage{bbding}

\usepackage{nomencl}
\makenomenclature
\usepackage{etoolbox}
\renewcommand\nomgroup[1]{%
  \item[\bfseries
  \ifstrequal{#1}{A}{Acronyms}{%
  \ifstrequal{#1}{S}{Symbols}{%
  \ifstrequal{#1}{U}{Units}{}}}%
]}

\NiceMatrixOptions{code-for-first-row = \scriptstyle,code-for-first-col = \scriptstyle }

\begin{document}

\title{A Differentially Private Energy Trading Mechanism Approaching Social Optimum}

\author{Yuji Cao,
Yue Chen,~\IEEEmembership{Member,~IEEE}
 %
\thanks{Y. Cao and Y. Chen are with the Department of Mechanical and Automation Engineering, The Chinese University of Hong Kong, Hong Kong SAR. (email: \{yjcao, yuechen\}@mae.cuhk.edu.hk) (\emph{Corresponding to Y. Chen})}
}

\markboth{Journal of \LaTeX\ Class Files,~Vol.~14, No.~8, September~2024}%
{Shell \MakeLowercase{\textit{et al.}}: Bare Demo of IEEEtran.cls for IEEE Journals}

\maketitle

\begin{abstract}
This paper proposes a differentially private energy trading mechanism for prosumers in peer-to-peer (P2P) markets, offering provable privacy guarantees while approaching the Nash equilibrium with nearly socially optimal efficiency.  We first model the P2P energy trading as a (generalized) Nash game and prove the vulnerability of traditional distributed algorithms to privacy attacks through an adversarial inference model. To address this challenge, we develop a privacy-preserving Nash equilibrium seeking algorithm incorporating carefully calibrated Laplacian noise. We prove that the proposed algorithm achieves $\epsilon$-differential privacy while converging in expectation to the Nash equilibrium with a suitable stepsize. Numerical experiments are conducted to evaluate the algorithm's robustness against privacy attacks, convergence behavior, and optimality compared to the non-private solution. Results demonstrate that our mechanism effectively protects prosumers' sensitive information while maintaining near-optimal market outcomes, offering a practical approach for privacy-preserving coordination in P2P markets.
\end{abstract}







       




\section{Introduction}

\subsection{Background}
\IEEEPARstart{T}{he} global transition towards low-carbon energy systems has led to the widespread adoption of distributed energy resources (DERs), such as rooftop photovoltaic (PV) panels and energy storage devices~\cite{akorede2010distributed}. This has given rise to a new class of energy stakeholders known as prosumers, who can flexibly manage their energy by altering their production and consumption patterns~\cite{mahmud2020internet}. However, challenges from the fluctuating generation and unpredictable demands also come along~\cite{khorasany2020new}. Peer-to-peer (P2P) energy trading has emerged as a promising solution to address these challenges. By allowing prosumers to trade with each other, P2P energy trading can potentially offer economic benefits to prosumers, while simultaneously aiding in balancing local power and energy~\cite{zhang2018peer}. The efficacy of P2P energy trading, nevertheless, hinges on well-designed mechanisms that can efficiently coordinate the participants with competing interests to achieve social optimum.

Game theory offers powerful tools for mechanism design and equilibrium analysis. Existing game-theoretic market mechanisms typically adopt frameworks such as Stackelberg games, cooperative games and generalized Nash games. The Stackelberg game model represents a hierarchical decision-making process, where the market operator first determines energy prices while anticipating prosumers' reactions and the prosumers follow to decide on their traded energy quantities~\cite{tushar2019grid}. Cooperative games, in contrast, focus on how prosumers can form coalitions to achieve mutually beneficial outcomes and then distribute the overall revenue~\cite{malik2022priority}. The revenue allocation is based on certain criteria such as Shapley value~\cite{mei2019coalitional} that consider the marginal contribution and Nash bargaining method~\cite{chen2023asymmetric}. Generalized Nash games, the focus of this paper, model P2P energy markets when prosumers make simultaneous decisions, each optimizing their own objective while considering the shared constraints~\cite{wang2021distributed}. This game effectively considers the interdependencies in constrained energy networks and enhances prosumers' flexibility, elevating their role from price-takers to active price-makers~\cite{chen2022energy}.

However, the aforementioned game-theoretic mechanisms often overlook a critical aspect: privacy protection. As demand-side resources become increasingly networked~\cite{sedhom2021iot}, they risk exposing sensitive energy data to cyber attacks~\cite{lin2012false}. To tackle the privacy concern, researchers have primarily explored three approaches: homomorphic encryption-based methods~\cite{fan2012somewhat}, differential privacy-based methods~\cite{mcsherry2007mechanism} and secret sharing methods~\cite{karnin1983secret}. Homomorphic-encryption-based approaches calculate on encrypted data, producing results identical to those from unencrypted data. However, their computational overhead is significant and only limited operations are supported~\cite{kogos2017fully}. Secret sharing methods divide the data into multiple shares distributed among parties, ensuring that individual shares do not reveal sensitive information. While ensuring individual share security, the reconstruction from distributed shares increases communication overhead and require careful coordination~\cite{karnin1983secret}. Alternatively, the \textit{differential privacy}-based method protects privacy by adding controlled noise to datasets~\cite{dwork2006differential}, preventing attackers from identifying individual data when querying on the overall dataset. Since differential privacy offers compatibility with arbitrary post-processing of the algorithm output, it has become a de facto tool for privacy preservation in recent years. However, implementing differential privacy in practical applications still faces challenges such as data correlation issue~\cite{han2015sensitive}, the trade-off between utility and privacy~\cite{hsu2014differential}, etc.

\begin{table*}[thbp]
\centering
\caption{Comparison with Existing Works Based on Differential Privacy}
\label{tab:related-work}
\resizebox{\textwidth}{!}{%
\begin{tabular}{c|cccccc}
\hline
Category & Ref. & Convergence Guarantee & Privacy Guarantee & Competitive & Strategy Set Interdependence & Adversarial Model \\ \hline
\multirow{2}{*}{Energy Trading} & \cite{wei2024peer} & $\times$ & $\checkmark$ & $\times$ & $\times$ & $\times$ \\
 & \cite{ahmed2024optimization} & $\times$ & $\times$ & $\times$ & $\times$ & $\checkmark$ \\ \hline
Distributed Optimization & \cite{huang2015differentially},\cite{han2016differentially},\cite{mo2016privacy},~\cite{nozari2016differentially},~\cite{nozari2017differentially} & $\checkmark$ & $\checkmark$ & $\times$ & $\times$ & $\times$ \\ \hline
\multirow{2}{*}{Game Theory} & \cite{wang2022differentially},\cite{ye2021differentially},\cite{wang2024differentially} & $\checkmark$ & $\checkmark$ & $\checkmark$ & $\times$ & $\times$ \\
 & \cite{wang2024ensuring},~\cite{wang2024quadratic} & $\checkmark$ & $\checkmark$ & $\checkmark$ & $\checkmark$ & $\times$ \\ \hline
\rowcolor[HTML]{EFEFEF}
Energy Trading \& Game Theory & Proposed & $\checkmark$ & $\checkmark$ & $\checkmark$ & $\checkmark$ & $\checkmark$ \\ \hline
\end{tabular}%
}
\end{table*}

In P2P energy trading, differential privacy-based methods remain largely unexplored, with only a few studies addressing this topic~\cite{wei2024peer},~\cite{ahmed2024optimization}. Since P2P market requires a coordinator to facilitate transactions by collecting and broadcasting aggregated market signals, the centralized collection of prosumer information introduces potential privacy vulnerabilities. To address this, Wei et al.~\cite{wei2024peer} adopted Laplace noise to perturb prosumers' gradient information before sharing. Similarly, another work~\cite{ahmed2024optimization} considered an additional aggregator between the utility grid and prosumers, adding Laplace noise to securely report aggregated prosumers' data to the utility grid. While these works incorporate differential privacy in P2P energy trading, the optimization problems are essentially cooperative, neglecting the competitive market dynamics or game-theoretic interactions among strategic prosumers. Moreover, the existing work relies on experimental validation through case studies, leaving a significant gap in rigorous theoretical analysis, particularly regarding convergence properties and privacy guarantees. This motivates us to bridge such a gap.
\subsection{Related Work}

Distributed Nash equilibrium seeking is an important field related to both game theory and consensus-based distributed optimization. In recent years, this field has been extensively investigated in network settings due to the proliferation of distributed systems such as smart grids. In game theory, equilibrium seeking has been studied in different games. For general network games, gossip-based algorithms~\cite{salehisadaghiani2016distributed} and inexact Alternating Direction Method of Multipliers (ADMM)~\cite{salehisadaghiani2019distributed} have been proposed. For non-general games such as aggregate games and non-cooperative multicluster games, consensus-based methods were used to estimate the aggregate decisions~\cite{koshal2016distributed} and clusters' objective functions~\cite{ye2019unified}, respectively. In consensus-based distributed optimization, gradient-based methods are gaining popularity for their computational efficiency and low storage requirements~\cite{nedic2009distributed},~\cite{qu2017harnessing}. For static consensus (or average consensus), researchers have combined gradient-descent with consensus operations on individual decision variables~\cite{nedic2009distributed}. To address the balanced-graph limitation, dynamic consensus approaches~\cite{yuan2016convergence} enable agents to track the global gradient, ensuring convergence at the cost of increased communication overhead. However, though algorithms in both fields demonstrate great results, the consensus-seeking process still requires iterative information sharing among players, potentially exposing private information.

Differential privacy has emerged as a key approach to protecting sensitive information in distributed systems. Researchers have applied this method to various distributed problems, such as optimization~\cite{huang2015differentially},~\cite{han2016differentially},~\cite{nozari2016differentially}, and average consensus\cite{mo2016privacy},~\cite{nozari2017differentially}. While these works relate to Nash equilibrium seeking, they differ fundamentally in agent behavior: distributed optimization involves cooperative agents working towards a common objective, whereas game-theoretic scenarios are competitive with self-interested players. In game theory, differential privacy has been extensively studied for aggregate games~\cite{wang2022differentially},\cite{ye2021differentially},\cite{wang2024differentially}. Some approaches perturb transmitted messages between agents~\cite{ye2021differentially},\cite{wang2024differentially}, providing convergence guarantees under specific conditions, such as globally bounded gradients and symmetric graphs. Alternatively, Wang et al.~\cite{wang2022differentially} perturbed gradients in optimization for stochastic aggregate games. However, these methods are inapplicable for generalized Nash equilibrium seeking---our more complex case---where payoffs depend on all individual actions, unlike aggregate games that consider only average actions. For generalized network games, Wang et al.\cite{wang2024ensuring} achieved provable convergence and differential privacy using diminishing step sizes, albeit at the cost of reduced convergence speed. Extending this line of research, another approach~\cite{wang2024quadratic} randomized projections to protect the influence graph. While these works provide valuable insights, they primarily focus on privacy preservation in non-adversarial settings. However, as privacy protection and attacks form an inter-evolving adversarial game, it is crucial to consider the potential adversary model before protection, even accounting for worst-case scenarios. A comparison of our work with the state-of-the-art works is provided in Table~\ref{tab:related-work}.
\subsection{Contribution and Organization}
In this paper, we propose a differentially private energy trading mechanism for prosumers in P2P markets, addressing the critical privacy issue in decentralized energy transactions. We formulate the problem as a generalized Nash game and propose an adversarial model to reveal privacy vulnerabilities under potential attacks. To counteract the attacks, we develop a privacy-preserving Nash equilibrium seeking algorithm with rigorous proofs of convergence and $\epsilon$-differential privacy guarantees. The main contributions are two-fold:
\begin{enumerate}
\item \textit{Privacy-Preserving Algorithm with Theoretical Guarantees:} We develop a novel privacy-preserving Nash equilibrium seeking algorithm for P2P energy trading. The algorithm incorporates carefully calibrated Laplacian noise to protect the sensitive demand information of prosumers. Through rigorous theoretical analysis, we prove that the proposed algorithm achieves $\epsilon$-differential privacy while maintaining convergence in expectation to the Nash equilibrium of the energy trading game. Additionally, we establish a theoretical bound on the variance of the deviation from the exact Nash equilibrium. Comprehensive numerical experiments are conducted to provide insights into the performance of the proposed algorithm under various privacy settings.
    \item \textit{Adversarial Inference Model:} We propose an adversarial inference model to demonstrate that the traditional distributed equilibrium seeking algorithm is vulnerable to privacy attacks. We prove that under the worst-case scenario, i.e., when an adversary has access to all but one prosumer's private information, it can infer the private information of the remaining prosumer accurately. Case studies demonstrate that the proposed privacy-preserving algorithm effectively resists adversarial inference attempts, even under the worst-case scenario. We analyze the required noise levels across different degrees of compromised information. Moreover, the convergence behavior and the quality of equilibrium are also investigated with discussions on the impacts of added noises on the prosumers' behaviors.

\end{enumerate}

The rest of this paper is organized as follows: The P2P energy sharing game and the traditional equilibrium seeking algorithm are introduced in Section \ref{sec:game}. An adversarial inference model is also presented in Section \ref{sec:game} to reveal that the traditional algorithm may lead to privacy leakage. To address this issue, a privacy-preserving equilibrium seeking algorithm is developed in Section \ref{sec:algorithm} with proven privacy and convergence properties. Case studies are conducted in Section \ref{sec:case} with conclusions drawn in Section \ref{sec:conclusion}.

\emph{Notations}: $\mathbb{R}^n$ denotes the set of $n$-dimensional real vectors and $\mathbb{R}^{m \times n}$ denotes the set of $m \times n$ real matrices. For a set $X$, $|X|$ denotes its cardinality. For a scalar $c$, $|c|$ denotes its absolute value. For a vector $x \in \mathbb{R}^n$ (a matrix $A \in \mathbb{R}^{m\times n}$), $x^T$ ($A^T$) denotes its transpose. We use $\mathbf{1}$ and $\mathbf{0}$ to denote vectors of ones and zeros, respectively. For $x,y \in \mathbb{R}^n$, we denote the inner product $x^T y = \sum_{i=1}^n x_i y_i$ where $x_i$, $y_i$ stand for the $i$-th entries of $x$ and $y$, respectively. For a vector $x$, $||x||$ denotes its Euclidean norm. $\mathbb{E}[\cdot]$ denotes the expectation operator and $\mathbb{P}[\cdot]$ denotes probability. For a random variable $\gamma$, $\gamma \sim \text{Lap}(0,2\sigma^2)$ means $\gamma$ follows a Laplace distribution with mean 0 and scale parameter $\sigma$. $\lim_{k \to \infty}$ denotes the limit as $k$ approaches infinity. For matrices, $\otimes$ denotes the Kronecker product. For a matrix $A$, $\text{Tr}(A)$ denotes its trace. For a vector $x = (x_1, ..., x_n)$, $\textbf{diag}(x_1, ..., x_n)$ denotes an $n \times n$ diagonal matrix with the elements of $x$ on its main diagonal. $\textbf{I}_n$ denotes the $n \times n$ identity matrix.
\section{Peer-to-peer Energy Trading Game}

\label{sec:game}
We consider a set of $I$ prosumers, indexed by $i \in \mathcal{I}=\{1,...,I\}$ in a standalone energy community. The electricity demand of each prosumer $i \in \mathcal{I}$ is $d_i$. To meet its demand, the prosumer $i \in \mathcal{I}$ can self-produce $p_i$ and buy $q_i$ from the P2P energy trading market. The cost of self-production is quantified by a quadratic function $c_ip_i^2$. Here, electricity demand $d_i$ is private information of prosumer $i \in \mathcal{I}$ that we aim to protect.

\subsection{Mathematical Model}
The intercept function bidding mechanism \cite{chen2020approaching} in our previous work is adopted for P2P energy trading. Specifically, the relationship between the traded energy $q_i$ and the trading price $\lambda$ of each prosumer $i \in \mathcal{I}$ is given by a generalized demand function:
\begin{align}\label{eq:demand-function}
    q_i = -a\lambda+b_i,
\end{align}
where $a>0$ is the market sensitivity. Each prosumer $i$ bids the intercept $b_i$ of the generalized demand function \eqref{eq:demand-function}. The prosumer $i$ buys from the P2P energy trading market if $q_i>0$; otherwise, the prosumer sells to the market. With all the bids from prosumers, the P2P energy trading price $\lambda$ is determined by the market clearing condition:
\begin{align}
    \sum \nolimits_{i \in \mathcal{I}} q_i = \sum \nolimits_{i \in \mathcal{I}} (-a\lambda+b_i) =0.
\end{align}

Based on the above settings, all prosumers in the P2P energy trading market play a generalized Nash game. Each prosumer $i \in \mathcal{I}$ solves:
\bsq \label{eq:prosumer}
\begin{align}
    \min_{p_i,b_i,q_i}~ & c_i p_i^2  + \lambda q_i, \label{eq:prosumer-1}\\
    \mbox{s.t.}~ & p_i + q_i = d_i, \label{eq:prosumer-2}\\
    ~ & q_i =-a\lambda +b_i, \label{eq:prosumer-3}\\
    ~ & \sum \nolimits_{i=1}^I q_i =0. \label{eq:prosumer-4}
\end{align}
\esq
The objective function \eqref{eq:prosumer-1} is to minimize the total cost, including the self-production cost plus the purchase cost from the P2P energy trading market. Constraint \eqref{eq:prosumer-2} indicates the energy balance of each prosumer. The energy trading market clearing condition is given by \eqref{eq:prosumer-4}, which is also a common constraint of all prosumers.

By using $b_i$ to represent $p_i$, $q_i$, and $\lambda$, the generalized Nash game \eqref{eq:prosumer} can be transformed into the equivalent Nash game below.  Each prosumer $i \in \mathcal{I}$ solves:
\begin{align}
    \max_{b_i}~ & - c_i \left(d_i+\frac{\sum_i b_i}{I}- b_i\right)^2  - \frac{\sum_i b_i}{Ia} \left(-\frac{\sum_i b_i}{I}+b_i\right),
\end{align}

By reorganizing the objective function and ignoring the constant terms, the decision-making problem for each prosumer $i \in \mathcal{I}$ is given by
\begin{align}
  \max_{b_i}~ &  -1/2 b_i^2 + \beta_i b_i + \sum_{j \in \mathcal{I}/\{i\}} \mu_{ij} b_i b_j,
\end{align}
where
\bsq\label{eq:Nash-game}
\begin{align}
    \beta_i=~ & \frac{ac_id_i I}{ac_i(I-1)+1}, \label{eq:beta}\\
    \mu_{ij}=~ & \frac{2ac_i(I-1) - (I-2)}{2(I-1)(ac_i(I-1)+1)}.
\end{align}
\esq

Denote the Nash game \eqref{eq:Nash-game} compactly as $\mathcal{G}=\{\mathcal{I}, \mathcal{B}, \Gamma\}$, where (1) $\mathcal{I}$ is the set of players (prosumers); (2) $\mathcal{B}:=\{\mathcal{B}_1,...,\mathcal{B}_I\}$ are the action sets of prosumers, with $\mathcal{B}_i = \{b_i \in \mathbb{R} \},\forall i \in \mathcal{I}$; (3) $\Gamma:=\{\Gamma_1,..., \Gamma_I\}$ are the payoff functions of prosumers, with $\Gamma_i (b_i, b_{-i}) = -1/2 b_i^2 + \beta_i b_i + \sum_{j \in \mathcal{I}/\{i\}} \mu_{ij} b_i b_j$. Here, $b_{-i}$ is a collection of all other prosumers' actions, i.e., $b_{-i}:=\{b_j, \forall j \ne i \}$.

\begin{definition}[Energy Trading Equilibrium]
   A profile $b^*$ is a Nash equilibrium of the P2P energy trading game $\mathcal{G}=\{\mathcal{I}, \mathcal{B}, \Gamma\}$ if and only if
    \begin{align}
        b_i^* \in \mbox{argmax}~ \Gamma_i(b_i, b_{-i}^*), ~\mbox{s.t.}~ b_i \in \mathcal{B}_i,~\forall i \in \mathcal{I}
    \end{align}
\end{definition}

\subsection{Traditional Nash Equilibrium Seeking Algorithm}
To achieve the energy trading equilibrium $b^*$, if without privacy considerations, we can apply the traditional Nash equilibrium seeking algorithm in \cite{shi2016network}, which is summarized in Algorithm \ref{algo-1}. Each prosumer $i \in \mathcal{I}$ maintains a local estimate of the bids of all prosumers, denoted by $y_i \in \mathbb{R}^I$. It can communicate with a set of neighbouring prosumers $N_{\text{com}}^i$ iteratively to update the estimate $y_i$. Let $\omega$ be the weight of the communication network that satisfies
$ 0 \le \omega \le \frac{1}{1+\max_{i \in \mathcal{I}} |N_{\text{com}}^i|}$. $\alpha$ is the step size and  $f_i:=e_i - [\mu_{i1}, ..., \mu_{iI}]^{\top}$, where $e_i$ is a unit vector whose entries are all zero except the $i$-th being one.

\begin{algorithm}[htbp]
\caption{Traditional Nash Equilibrium Seeking}\label{algo-1}
\begin{algorithmic}[1]
\renewcommand{\algorithmicrequire}{ \textit{\textbf{Initialization:}}}
\REQUIRE
\STATE{Prosumer parameters $c_i, d_i, \forall i \in \mathcal{I}$, step size $\alpha$, weight of communication network $\omega$, iteration index $k=0$, and $y_i(0) \in \mathbb{R}^I, \forall i \in \mathcal{I}$, error tolerance $\tau$.}
\renewcommand{\algorithmicrequire}{ \textit{\textbf{Output:}}}
\REQUIRE
\STATE{$y_i(k)$ at each node $i \in \mathcal{I}$.}
\renewcommand{\algorithmicrequire}{ \textit{\textbf{Equilibrium Seeking:}}}
\REQUIRE
\WHILE{$\sum_i||y_i(k+1)-y_i(k)|| \ge \tau$}
\FOR{Each prosumer $i \in \mathcal{I}$}
\STATE{Sends $y_i(k)$ to its neighbours in the set $N_{\text{com}}^i$.}
\STATE{Updates its state according to}
\begin{align}
    y_i(k+1)=~ & y_i(k)-\omega \sum_{j \in N_{\text{com}}^i} [y_i(k) - y_j(k)] \nonumber\\
    ~ & - \alpha f_i [f_i^{\top} y_i(k) - \beta_i]
\end{align}
\ENDFOR
\STATE {Set $k=k+1$}
\ENDWHILE
\end{algorithmic}
\end{algorithm}

The traditional Nash equilibrium seeking algorithm (Algorithm \ref{algo-1}) has proven convergence and optimality guarantees (i.e., $\lim_{k \to \infty} y_i(k)=b^*,\forall i$) and has been widely used in the coordination of demand response resources. However, as seen in the next section, an adversary is able to reverse-engineer the private information of prosumers (the electricity demand $d_i$) from the coordination variables $y_i(k)$ exchanged among prosumers during iterations.

\subsection{Adversarial Inference Model}
In this section, we propose an adversarial inference model for privacy attack on the energy sharing game under the \emph{worst-case scenario}: An adversary obtains the private information of all but one prosumer, and aims to infer the private information of the remaining prosumer. Take the privacy attack on prosumer $i\in \mathcal{I}$ as an example. Suppose the adversary has access to the values of coordination variable $y_{i}(k)$ during the iterations $k=\hat{k}_1,...,\hat{k}_2$ and the private coefficient $\beta_j,\forall j \in \mathcal{I}/\{i\}$ of all prosumers except $i$. Then, the adversarial model for inferring the private information $\hat{\beta}_{i}$ of prosumer $i$ is formulated as follows:
\bsq\label{eq:infer}
\begin{align}
    \min_{\hat{\beta}_{i}, z_l(k),\forall k, \forall l \in \mathcal{I}}~ & \sum \nolimits_{k=\hat{k}_1}^{\hat{k}_2} \left( z_{i}(k)-\bar{y}_{i}(k)\right)^2,\label{eq:infer-1} \\
    \mbox{s.t.}~ & z_{i}(\hat{k}_1)=\bar{y}_{i}(\hat{k}_1), \label{eq:infer-2}\\
    ~ & z_l(k+1) = z_l(k) - \omega \sum_{j \in N_{\text{com}}^l} [z_l(k)-z_j(k)] \nonumber\\
    ~ & - \alpha f_l[f_l^{\top} z_l(k) -\hat{\beta}_l], \forall l \in \mathcal{I}, \label{eq:infer-3}\\
    ~ & \hat{\beta}_l = \beta_l, \forall l \in \mathcal{I} \backslash \{i\}.  \label{eq:infer-4}
\end{align}
\esq

Here, $\hat{\beta}_{i}$ is the inferred value of $\beta_{i}$, which is a decision variable. With $\hat{\beta}_{i}$, the inferred $\hat{d}_{i}$ can be determined uniquely by \eqref{eq:beta}. 
The objective function \eqref{eq:infer-1} is to minimize the Euclidean distance between the coordination variable trajectory of prosumer $i$ with the inferred $\hat{\beta}_{i}$ (i.e., $\{z_{i}(k),\forall k=\hat{k}_1,...,\hat{k}_2\}$) and the actual coordination variable trajectory (i.e., $\{\bar y_{i}(k),\forall k=\hat{k}_1,...,\hat{k}_2\}$). Constraint \eqref{eq:infer-2} initializes the values of coordination variables. The iterations are given in \eqref{eq:infer-3}, and \eqref{eq:infer-4} means that the adversary knows the private information of all prosumers but $i$.

The following proposition indicates that the accurate $\beta_i$ can be inferred using the adversarial inference model \eqref{eq:infer}.

\begin{proposition}
\label{thm-adversarial}
In a peer-to-peer energy trading game $\mathcal{G}=\{\mathcal{I}, \mathcal{B}, \Gamma\}$ with a fully connected communication network, suppose the adversary has access to
\begin{enumerate}
\item The coordination variable values $y_{i}(k)$ for prosumer $i$ during iterations $k = \hat{k}_1, ..., \hat{k}_2$ with $\hat{k}_2 - \hat{k}_1 \ge |\mathcal{I}|$;
\item The private coefficients $\beta_j$ for all other prosumers $j \in \mathcal{I} \backslash \{i\}$.
\end{enumerate}
Then, the private coefficient $\beta_i$ (and the corresponding $d_i$) of prosumer $i$ can be accurately inferred by the adversarial inference model \eqref{eq:infer}.


\end{proposition}

The proof of Proposition \ref{thm-adversarial} can be found in Appendix \ref{appendix:thm-adversarial}. The following example illustrates that even when $\hat{k}_2 - \hat{k}_1 < |\mathcal{I}|$, the traditional Nash equilibrium seeking algorithm may still suffer from privacy attack.
\begin{table}[!htbp]
  \centering
  \caption{Parameters of Prosumers}\label{tab:parameter}
  \begin{tabular}{ccccccc}
  \hline
    Prosumer  & 1    & 2     &  3   &  4    &  5   &  6 \\
    \hline
    $c_i$ (\$/kWh$^2$) &  0.015 &  0.03 &  0.02 & 0.015 & 0.025 & 0.03 \\
    $d_i$ (kWh) & 15 &  18 &  25 & 20 & 18 & 20 \\
    $\beta_i$ (kWh) & 15.88 & 20.25 & 27.27 & 21.18 & 20.00 & 22.50\\
    \hline
   \end{tabular}
\end{table}

\begin{figure}[!htbp]
  \centering
  \includegraphics[width=0.45\textwidth]{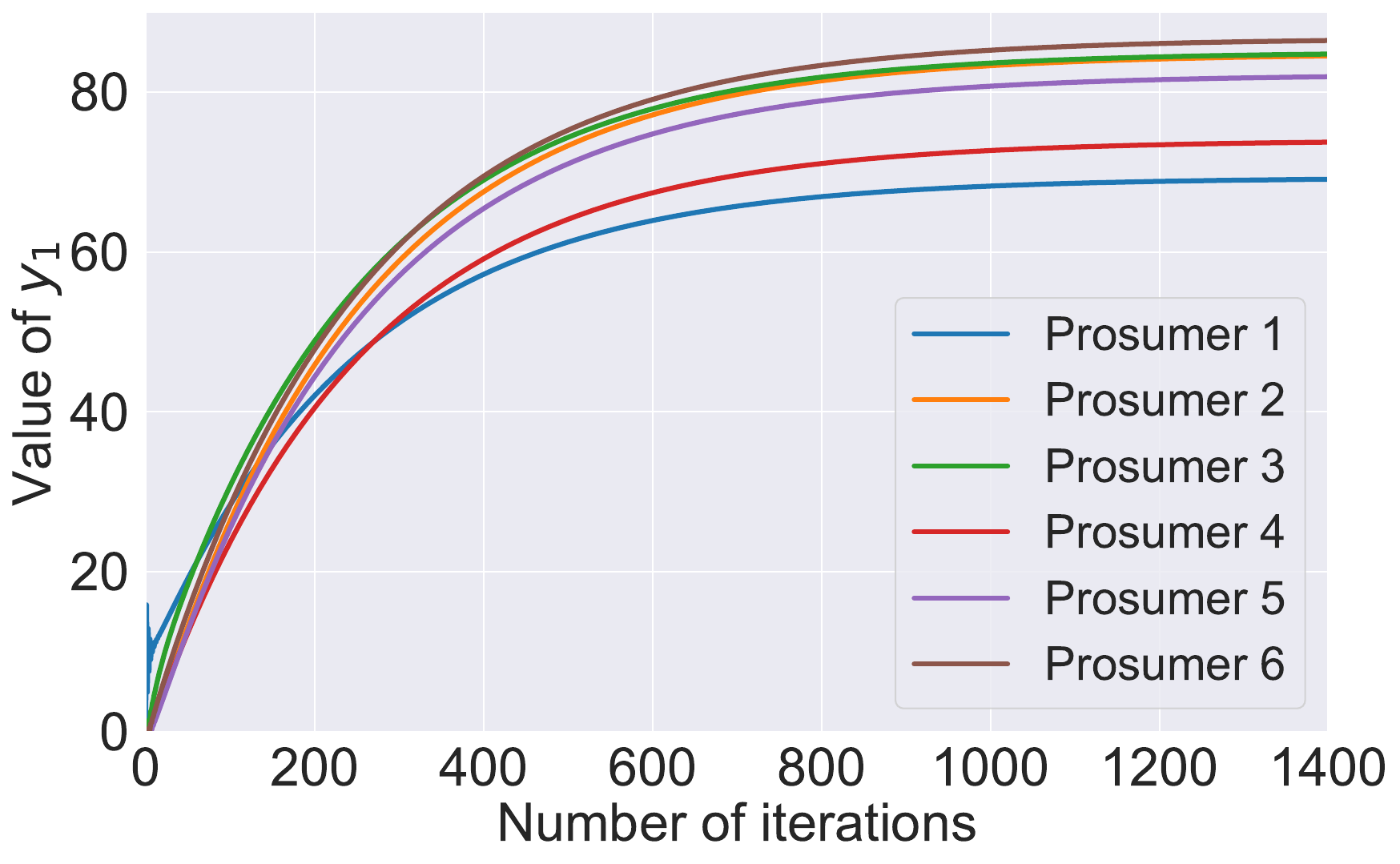}\\
  \caption{Change of $y_1$ during iterations}\label{fig:algo-1}
\end{figure}

\emph{Example}: Suppose there are 6 prosumers participating in the P2P energy trading market, whose parameters are provided in Table \ref{tab:parameter} and the market sensitivity $a=100$ kWh/\$. The corresponding $\beta_i$ calculated by \eqref{eq:beta} is also given in the table. Suppose the adversary already knows the private information $d_i$ of prosumers 2-6 and wants to obtain the private information (demand) of prosumer 1. First, we apply Algorithm 1 to derive the Nash equilibrium. The change of $y_1$ during iterations is shown in Fig. \ref{fig:algo-1}. We can see that Algorithm \ref{algo-1} converges to $[69.28, 84.77, 85.00, 73.96, 82.17, 86.71]$ kWh. By checking the best responses of prosumers of the game $\mathcal{G}=\{\mathcal{I},\mathcal{B},\Gamma\}$, we can easily confirm that Algorithm \ref{algo-1} converges to the energy trading equilibrium $b^*$. However, we also find that Algorithm \ref{algo-1} may lead to privacy leakage. To be specific, we randomly select sets of iterations $\{\hat{k}_1,...,\hat{k}_2\}$ and apply the adversarial inference model to infer the private information of prosumer 1. The results are shown in Table \ref{tab:adverse}. We can see that the proposed adversarial inference model can 100\% get the actual value of the private demand $d_1$ of prosumer 1 even only with information about very few iterations.

\begin{table}[!htbp]
  \centering
  \caption{Inferred value of $\beta_1$ and $c_1$}\label{tab:adverse}
  \begin{tabular}{ccccccc}
  \hline
    $\hat{k}_1$  & 1    &  12    &  23  &  27    & 100\\
    $\hat{k}_2$ &  5  & 16 & 26 & 30  & 102  \\
    No. of known iterations & 5 & 5 & 4 & 4 & 3 \\
    \hline
    $\hat{\beta}_1$ (kWh) & 15.88 & 15.88 & 15.88 & 15.88 & 15.88\\
    $\hat{d}_1$ (kWh) &  15 &  15 & 15  & 15 & 15\\
    \hline
   \end{tabular}
\end{table}

\section{Privacy-preserving Nash Equilibrium Seeking}
\label{sec:algorithm}
To address the privacy concern, in the following, we propose a privacy-preserving Nash equilibrium seeking algorithm. The algorithm is proven to be $\varepsilon$-differential private and converge in expectation to the Nash equilibrium of the game $\mathcal{G}=\{\mathcal{I},\mathcal{B},\Gamma\}$.

\subsection{Algorithm Design}
Before going into the details of the proposed algorithm, we first introduce some concepts related to differential privacy.

\begin{definition}[$\mu$-adjacent]
The datasets $d:=\{d_i,\forall i \in \mathcal{I}\}$ and $d':=\{d_i',\forall i \in \mathcal{I}\}$ are $\mu$-adjacent if they only differ in one element by $\mu$, ,
\begin{align}
    \exists i, \mbox{s.t.}~ ||d_i- d_i'||_1 \le \mu ~\mbox{and}~ d_j=d_j',\forall j \ne i.
\end{align}
\end{definition}

The concept ``$\mu$-adjacent'' is used to describe the similarity between two datasets, while the concept ``$\varepsilon$-differential privacy'' describes a situation that when two datasets are similar, their outputs are non-differentiable. Let $\mathcal{M}$ denote a Nash equilibrium seeking algorithm that maps the private demand $d$ to the prosumers' bids $b_i,\forall i$.


\begin{definition}[$\varepsilon$-differential privacy \cite{dwork2014algorithmic}]
A randomized equilibrium seeking algorithm $\mathcal{M}: \mathbb{R}_{+}^{I} \to \mathbb{R}^{I}$ from the demand $d$ to optimal bids $b$ is $\varepsilon$-differentially private if for all $\mathcal{S} \subseteq \mbox{Range}(\mathcal{M})$ and for any pair $(d, d')$ that are $\mu$-adjacent:
\begin{align}
    \mathbb{P}[\mathcal{M}(d) \in \mathcal{S}] \le \text{exp}(\varepsilon) \mathbb{P}[\mathcal{M}(d') \in \mathcal{S}].
\end{align}
\end{definition}

In the following, we introduce the proposed privacy-preserving Nash equilibrium seeking algorithm based on the Laplace mechanism. Particularly, a Laplacian noise $\gamma$ that follows the Laplace Distribution is included in the iterations. $\text{Lap}(0,2\sigma_{\gamma}^2)$ refers to the Laplace distribution with probability density function:
\begin{align}
    \text{Lap}(0,2\sigma_{\gamma}^2) = \frac{1}{2\sigma_{\gamma}} \text{exp} \left(-\frac{|\gamma|}{\sigma_{\gamma}}\right)
\end{align}

The proposed privacy-preserving Nash equilibrium seeking algorithm is presented in Algorithm \ref{algo-2}. The key difference between Algorithms \ref{algo-2} and \ref{algo-1} lies in the equation \eqref{eq:update-privacy}. Instead of updating the state $y_i(k+1)$ using the accurate $\beta_i$, a Laplacian noise $\gamma_i$ is added. In the next section, we prove that with such a design, the proposed algorithm can achieve a good balance between privacy and optimality.

\begin{algorithm}[bhtp]
\caption{Privacy-preserving Equilibrium Seeking}\label{algo-2}
\begin{algorithmic}[1]
\renewcommand{\algorithmicrequire}{ \textit{\textbf{Initialization:}}}
\REQUIRE
\STATE{Prosumer parameters $c_i, d_i, \forall i \in \mathcal{I}$, step size $\alpha$, weight of communication network $\omega$, iteration index $k=0$, and $y_i(0) \in \mathbb{R}^I$, error tolerance $\tau$.}
\STATE{Each prosumer $i$ generates an i.i.d. Laplacian noise $\gamma_i\sim\text{Lap}(0, 2\sigma_{\gamma}^2)$.}
\renewcommand{\algorithmicrequire}{ \textit{\textbf{Output:}}}
\REQUIRE
\STATE{$y_i(k)$ at each node $i \in \mathcal{I}$.}
\renewcommand{\algorithmicrequire}{ \textit{\textbf{Equilibrium Seeking:}}}
\REQUIRE
\WHILE{$\sum_i||y_i(k+1)-y_i(k)|| \ge \tau$}
\FOR{Each prosumer $i \in \mathcal{I}$}
\STATE{Sends $y_i(k)$ to its neighbours in the set $N_{com}^i$.}
\STATE{Updates its state according to}
\begin{align}
    y_i(k+1)=~ & y_i(k)-\omega \sum_{j \in N_{com}^i} [y_i(k) - y_j(k)] \nonumber\\
    ~ & - \alpha f_i [f_i^{\top} y_i(k) - (\beta_i + \gamma_i)] \label{eq:update-privacy}
\end{align}
\ENDFOR
\STATE {Set $k=k+1$}
\ENDWHILE
\end{algorithmic}
\end{algorithm}

\subsection{Proof of Properties}
First, we prove that the proposed algorithm (Algorithm \ref{algo-2}) can preserve privacy by the following theorems.

\begin{theorem}
\label{thm-1}
For any pair $(d,d')$ that are $\mu$-adjacent, if $\sigma_\gamma \ge A\mu/\varepsilon$, where $A=\max\{ac_iI/(ac_i(I-1)+1),\forall i\}$, then Algorithm \ref{algo-2} is $\varepsilon$-differential private for any finite number of iterations $K$.
\end{theorem}

The proof of Theorem \ref{thm-1} can be found in Appendix \ref{appendix:thm-1}. It shows that the proposed Algorithm \ref{algo-2} can protect the privacy of prosumers. However, adding a random noise to the iteration processes may sacrifice optimality. Hence, another issue that we care about is whether the proposed algorithm can still converge to the Nash equilibrium $b^*$ of the energy trading game. We address this issue in Theorem \ref{thm-2} below.

\begin{figure*}[htbp]
  \centering
  \includegraphics[width=0.95\textwidth]{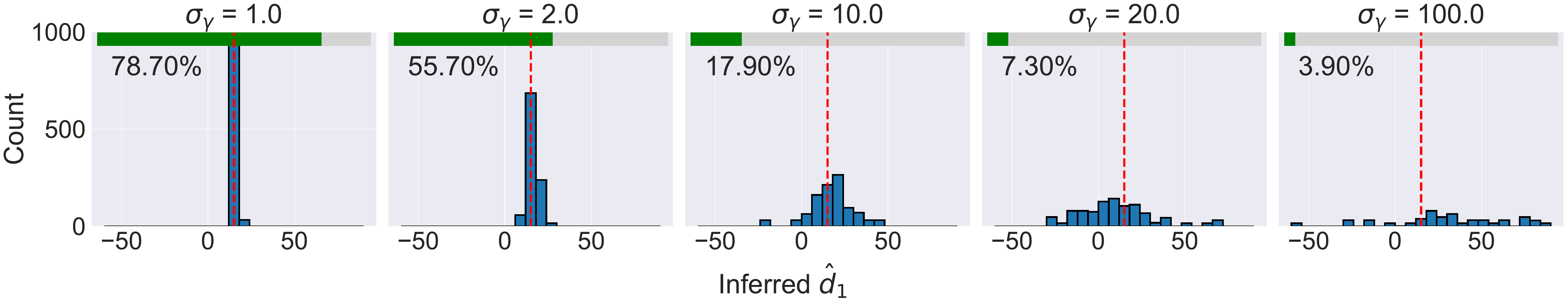}
  \caption{Distribution of inferred $\hat{d}_1$ of prosumer 1 under different noise levels when attack budget $B = 4$. The red vertical line indicates the true $d_1$ value $15$. The top green progress bar indicates the percentage of data falling within $\pm10$\% of the true value.}\label{fig:inferred-b-noise}
\end{figure*}

\begin{figure*}[hbt]
  \centering
  \includegraphics[width=0.95\textwidth]{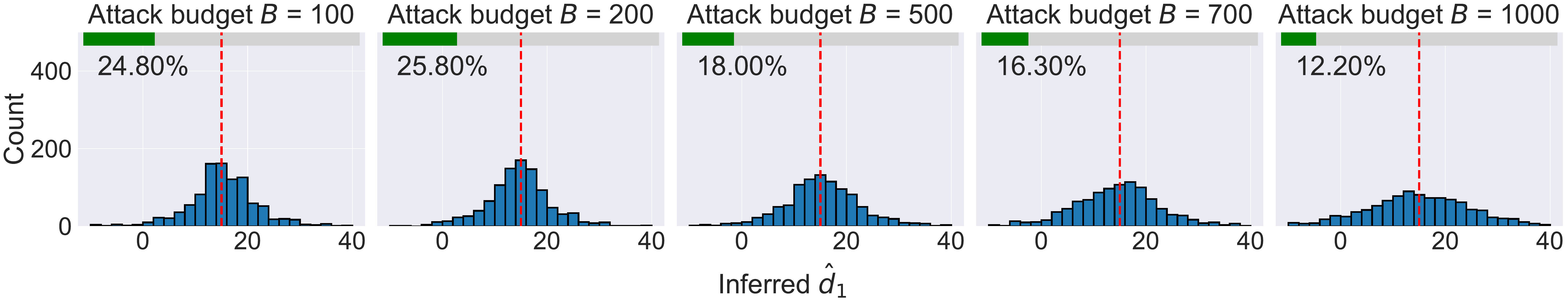}
  \caption{Distribution of inferred $\hat{d}_1$ of prosumer 1 under different attack budgets when noise level $\sigma = 5$. The red vertical line indicates the true ${d}_1$ value $15$. The top green progress bar indicates the percentage of data falling within $\pm10$\% of the true value.}\label{fig:inferred-b-attack}
\end{figure*}

\emph{A1}:
The step size $\alpha$ in Algorithm \ref{algo-2} satisfies
\begin{align}
0 < \alpha <  \min\{\frac{2-\overline{\lambda}}{3\overline{f}^2}, \frac{\underline{\lambda} \underline{f}^2}{2(\overline{f}^4 - \underline{f}^4)}\},
\end{align}
where
\begin{align}
    \overline{f} = ~ & \max\{||f_1||,...,||f_I||\}, \\
    \underline{f} = ~ & \min\{||f_1||,...,||f_I||\}, \\
    \overline{\lambda} = ~ & \max\{\lambda_2, ..., \lambda_I\}, \\
    \underline{\lambda} = ~ & \min\{\lambda_2, ..., \lambda_I\},
\end{align}
and $\lambda_1, \lambda_2, ..., \lambda_I$ are the eigenvalues of matrix $\tilde{\mathcal{W}}$ defined in Appendix \ref{appendix:thm-1} with $\lambda_1=0$.

\begin{theorem}\label{thm-2}
When assumption A1 holds, we have $\forall i \in \mathcal{I}$
\begin{align}
    \lim_{k \to \infty} \mathbb{E}[y_i(k)] = ~ & b^*, \\
    \lim_{k \to \infty} \mathbb{E}\|y_i(k) - b^*\|^2 \leq ~ &  \frac{2\alpha^2 I  \sigma_{\gamma}^2 \max\{||f_i||^2,\forall i\}}{1-m^2},
\end{align}
where $m$ is the spectral radius of matrix $Q$ defined in \eqref{matrix:Q}.
\end{theorem}

The proof of Theorem \ref{thm-2} can be found in Appendix \ref{appendix:thm-2}. It indicates that the proposed algorithm can not only protect privacy but also converge in expectation to the Nash equilibrium of the P2P energy trading game.

\section{Case Studies}
\label{sec:case}

In this section, numerical experiments are conducted to validate the effectiveness of the proposed algorithm. First, we assess the ability of the proposed algorithm to protect sensitive information from privacy attacks. Then, the convergence of the proposed algorithm under different noise scales is tested. Furthermore,
a fidelity analysis is performed to quantify the efficiency losses and the deviations from the Nash equilibrium, showing the trade-off between privacy and optimality. All optimization problems are modeled in MATLAB and solved using the GUROBI solver, with simulations run on a laptop with an Apple Silicon M3 Max Chip and 48 GB RAM. 

We consider an energy community with $\lvert \mathcal{I} \rvert = 6$ prosumers connected with each other. Without loss of generality, the weight of communication graph is uniformly set as $\omega = 0.1$; the electricity demand (private information) is $\boldsymbol{d} = [15; 18; 25; 20; 18; 20]$ kWh; the market sensitivity is set as $a = 100$ kWh/\$; the cost coefficient of self-production is set as $\boldsymbol{c} = [0.015; 0.03; 0.02; 0.015; 0.025; 0.03]$ \$/kWh$^2$. Given the above information, $\beta_i$ for each prosumer can be calculated with \eqref{eq:beta}. The step size $\alpha$ and initial estimate $y(0) \in \mathbb{R}^{I \times I}$ are set as $0.4$ and all-zero vectors, respectively.

\begin{figure}[!btp]
  \centering
  \includegraphics[width=0.4\textwidth]{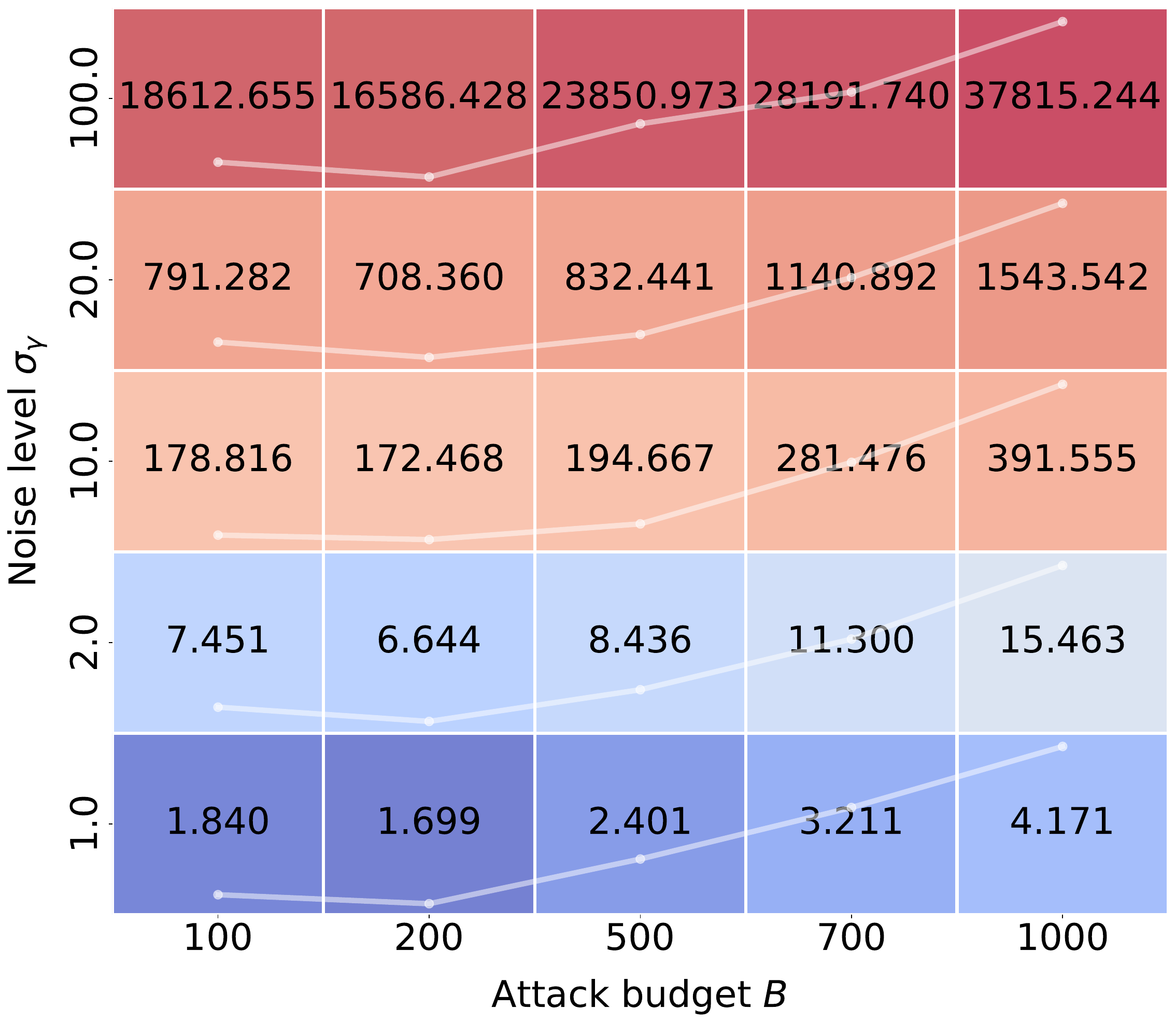}
  \caption{Heatmap of mean squared inference error.}\label{fig:heatmap}
\end{figure}

\begin{figure*}[!htbp]
  \centering
  \includegraphics[width=0.95\textwidth]{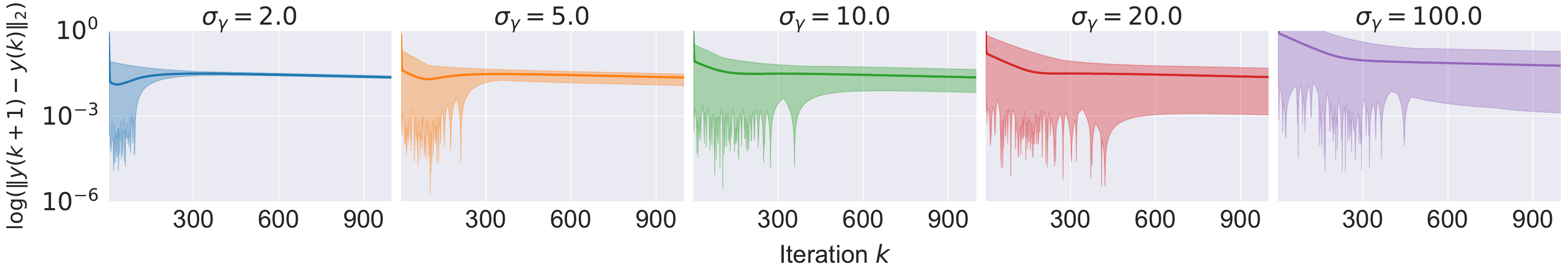}
  \caption{Convergence of Algorithm~\ref{algo-2} under various perturbations, with the $y$-axis on a logarithmic scale. The solid lines represent the average residual norm, and shaded areas show the variation of the residual norm.}\label{fig:cvg}
\end{figure*}

\begin{figure}[!htbp]
  \centering
  \includegraphics[width=0.4\textwidth]{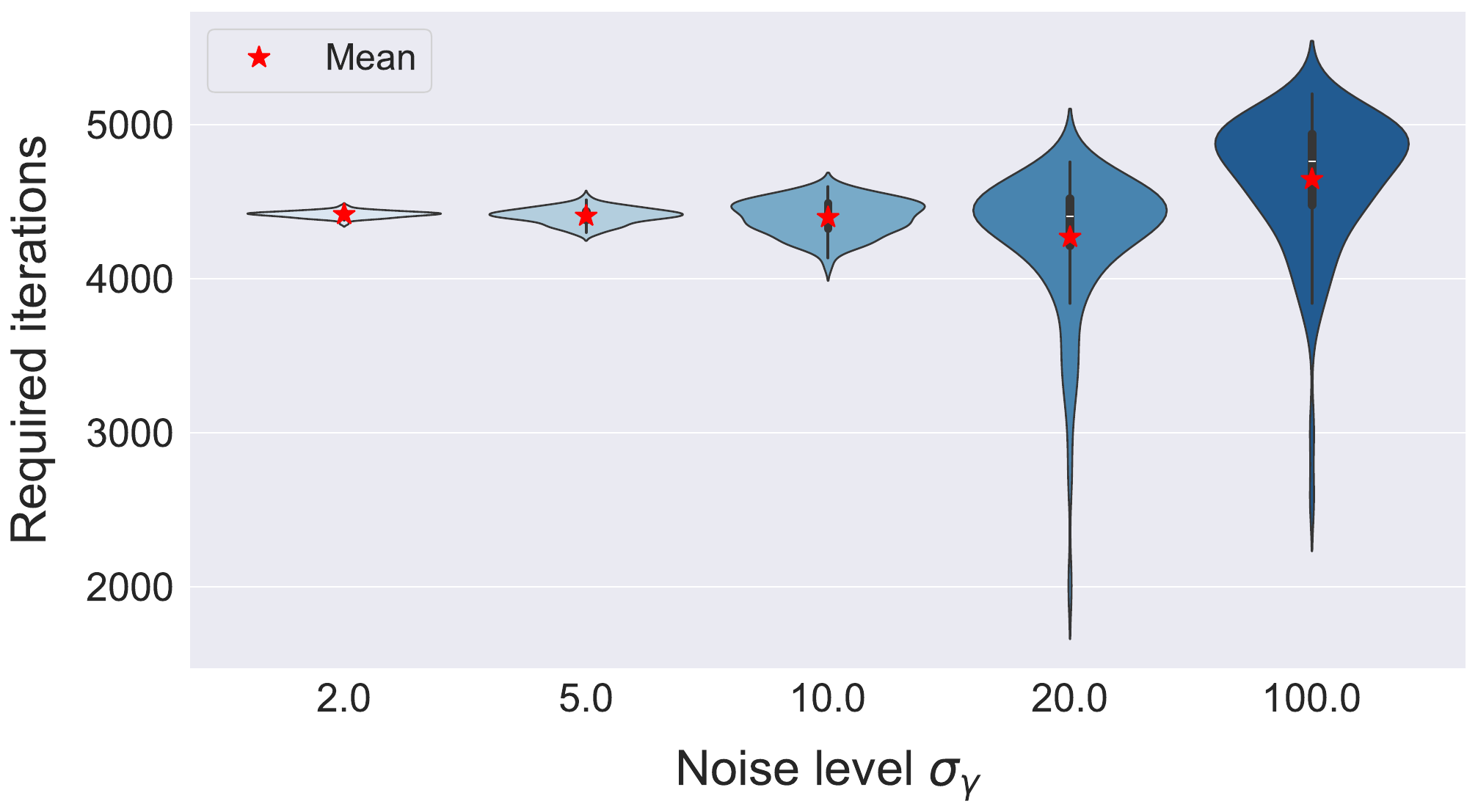}
  \caption{Distribution of required iterations for convergence under different noise levels.}\label{fig:violin}
\end{figure}

\subsection{Privacy Guarantee}

We assess the ability of the proposed algorithm to protect privacy through two experiments using the adversarial inference model \eqref{eq:infer}. Suppose the attacker wants to obtain the private demand information $d_1$ of prosumer 1. In the first experiment, the attack budget $B$ (the number of known iterations $\hat{k}_2-\hat{k}_1+1$) is fixed and the noise level varies. In the second experiment, the noise level is fixed and the attack budget varies. In both experiments, Algorithm \ref{algo-2} and the corresponding attack are run $1000$ times to ensure statistical significance. The distributions of the inferred $\hat{d}_1$ are recorded in Fig. \ref{fig:inferred-b-noise} and Fig. \ref{fig:inferred-b-attack}, respectively.

The first experiment provides insight into the noise level needed to protect sensitive information from accurate inference. We execute Algorithm~\ref{algo-2} with the noise level $\sigma_{\gamma}$ being set to $1$, $2$, $10$, $20$, and $100$, respectively. The larger the $\sigma_{\gamma}$, the larger the noise. We fix the attack budget at $4$ and initiate the attack from iteration $100$. The results of this experiment are provided in Fig.~\ref{fig:inferred-b-noise}. The percentage of inferred $\hat{d}_1$ that falls within $\pm10$\% of the actual value is also given. We can see that with an increasing $\sigma_{\gamma}$, the inferred $\hat{d}_1$ deviates more substantially from the true value of $15$, and the probability of recovering the actual private demand information reduces.

The second experiment explores the performance of the proposed algorithm under varying attacker capabilities. We set the attack budget $B$ (the number of known iterations $\hat{k}_2-\hat{k}_1+1$) to $100$, $200$, $500$, $700$, and $1000$, respectively.  The noise level $\sigma_{\gamma}$ is fixed at $5$ and the privacy attacks start at the 100th iteration. The simulation results are presented in Fig.~\ref{fig:inferred-b-attack}. Intuitively, one may expect the attacker's inference accuracy should improve with access to more iterations. However, contrary to this expectation, we observe a non-monotonic trend in the inference accuracy as the attack budget increases. The percentage of ``accurate'' $\hat{d}_1$ (those falling within $\pm10$\% of the actual value) first increases (from $24.80$\% to $25.80$\%) and then decreases (from $25.80$\% to $12.20$\%). This is probably caused by the accumulation of noise over iterations. With more iterations exposed to the attacker, on the one hand, it gains more data for inferring the private information; on the other hand, a higher level of noise is included in the data, leading to lower inference accuracy.

To see if this compounded effect exists for more general cases, we calculate the Mean Squared Error (MSE) of the inferred $\hat{d}_1$ under different noise level $\sigma_{\gamma}$ and attack budget $B$. The results are recorded in Fig.~\ref{fig:heatmap}. We can find that, under each given $\sigma_{\gamma}$, the MSE always first drops and then rises, as the attack budget $B$ grows. This validates that a proper introduction of random noises can effectively obscure private information, even against attackers obtaining access to data of more iterations. Moreover, under each attack budget $B$, the MSE always increases with a rising $\sigma_{\gamma}$, which is similar to the findings in the first experiment.

\subsection{Convergence Analysis}
Apart from the ability to protect private information, whether the proposed algorithm can still converge is also important. Here, we test the convergence of the proposed algorithm under different $\sigma_{\gamma}$ (noise scale). $\sigma_{\gamma}$ is set as $2$, $5$, $10$, $20$, and $100$, respectively. We run Algorithm~\ref{algo-2} for $100$ times under each $\sigma_{\gamma}$ with the step size $\alpha=0.05$. The evolution of residuals (quantified by $\text{log}||y(k+1)-y(k)||_2$) is shown in Fig.~\ref{fig:cvg}. The solid line shows the average situation while the shaded area shows the variation.  We can see from Fig.~\ref{fig:cvg} (from left to right) that the slope of the solid line decreases as $\sigma_{\gamma}$ increases. This indicates a slower convergence rate under an increasing noise level. Moreover, the shaded area widens as more noise is introduced, meaning that the iteration process becomes more uncertain. Still, in all our test cases, Algorithm \ref{algo-2} always converges.

In addition, the number of iterations needed for convergence under the error tolerance of $\tau=10^{-5}$ is recorded in Fig.~\ref{fig:violin}. The violin plots reveal that as $\sigma_\gamma$ decreases from $100$ to $2$, the distribution of required iterations becomes more concentrated. Moreover, the average number of iterations required for convergence under these five scenarios are $[4646.64, 4270.68, 4399.24, 4408.42, 4418.99]$, respectively, which are very close. This suggests that while larger $\sigma_\gamma$ leads to a slower convergence rate initially, the average convergence performance is relatively consistent across different noise scales.

\begin{figure*}[hbt]
  \centering
  \includegraphics[width=\textwidth]{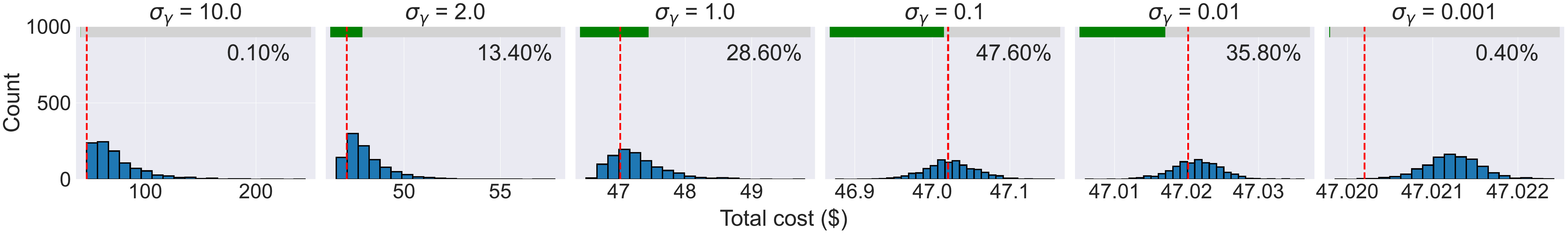}
  \caption{Distribution of total cost in Algorithm~\ref{algo-2} under different noise levels when market sensitivity $a$ set as $10$. The red vertical line indicates the total cost of Algorithm~\ref{algo-1}, while the top green progress bar indicates the percentage of samples with negative costs.}\label{fig:total-cost-dist}
\end{figure*}

\begin{figure}[bhtp]
  \centering
  \includegraphics[width=0.45\textwidth]{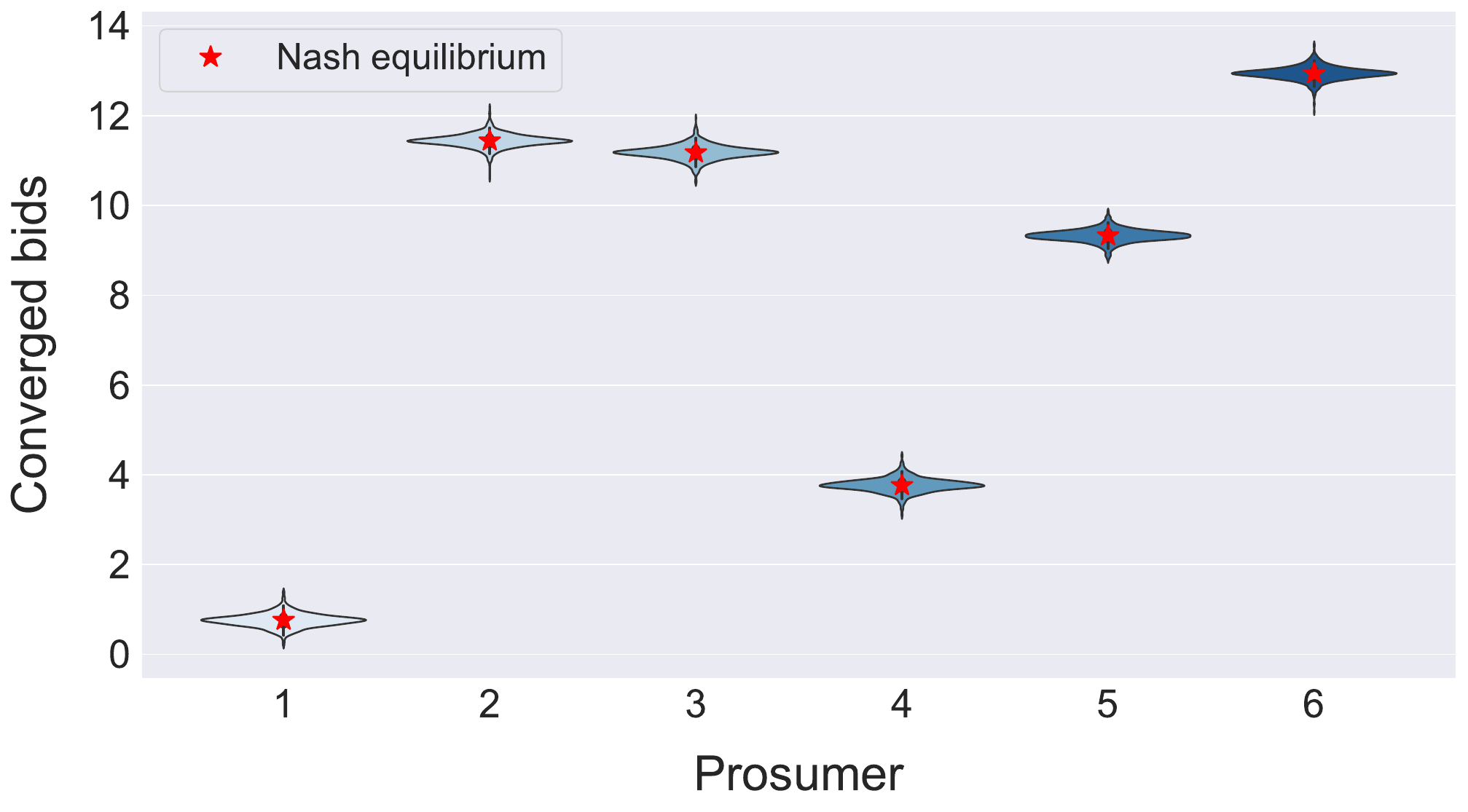}
  \caption{Distribution of converged bidding values for all prosumers when $\sigma_{\gamma} = 0.1$.}\label{fig:fidelity-violin}
\end{figure}

\subsection{Fidelity Analysis}
The above section validates the convergence of the proposed algorithm, while in this section we evaluate the quality of the converged results through the fidelity analysis. The quality is assessed from two aspects: (1) \emph{Average Cost Gap}: the average gap between the total cost $\sum_i c_i p_i^2$ under Algorithms \ref{algo-1} and \ref{algo-2}. (2) \emph{Equilibrium Deviation}: The gap between the converged result and the Nash equilibrium $b^*$.

We first test the impact of noise level (or $\sigma_\gamma$) on the total cost when the market sensitivity $a$ is set as $10$. The total costs under the proposed algorithm and the Nash equilibrium $b^*$ are summarized in Fig. \ref{fig:total-cost-dist}. As $\sigma_\gamma$ decreases from $10$ to $0.001$, we observe a significant reduction in the average cost gap. Notably, the gap decreases from $27.558$ at $\sigma_\gamma = 10$ to almost $0$ when $\sigma_\gamma \ge 0.1$, indicating that smaller $\sigma_{\gamma}$ (lower noise) leads to solutions closer to the Nash equilibrium. Furthermore, we find that the deviation due to the added noise may offset the negative impact of strategic prosumer behaviors, leading to a lower total cost than that at the Nash equilibrium. In Fig. \ref{fig:total-cost-dist}, for the cases with $\sigma_{\gamma}=2, 1, 0.1, 0.01$, more than 10\% of the samples have a lower total cost. To examine if this phenomenon occurs consistently, we repeat the experiments under varying market sensitivities. The results are presented in Table~\ref{tab:optimality-loss}. Under different market sensitivities, the percentage of samples with a negative cost gap consistently follows a pattern of initial increase followed by a decrease, which confirms that the noise effect exists in general. This is because when $\sigma_{\gamma}$ is large, the noise dominates and the prosumers choose their bids almost randomly, which results in high total costs; as noise reduces, the prosumers may converge to situations with lower total costs thanks to the randomness; when noise is extremely small, the strategic behaviors dominate and the total costs nearly equal that under the Nash equilibrium.

\begin{table}[bthp]
\centering
\begin{threeparttable}
\caption{Cost Gap under Different $\sigma_{\gamma}$ and Market Sensitivity $a$}
\label{tab:optimality-loss}
\begin{tabular}{c|ccc|ccc}
\hline
\multirow{2}{*}{\textbf{$\sigma_{\gamma}$}} & \multicolumn{3}{c|}{Average cost gap (\$)} & \multicolumn{3}{c}{Negative cost gap (\%)\tnote{1}} \\ \cline{2-7} 
      & $a$ = 10 & $a$ = 20 & $a$ = 50 & $a$ = 10 & $a$ = 20 & $a$ = 50 \\ \hline
10    & 27.558   & 20.810   & 19.010   & 0.1\%    & 0.0\%    & 0.0\%    \\
2     & 1.066    & 0.864    & 0.748    & 13.4\%   & 7.3\%    & 1.0\%    \\
1     & 0.264    & 0.223    & 0.185    & 28.6\%   & 19.4\%   & 4.6\%    \\
0.1   & 0.002    & 0.002    & 0.002    & 47.6\%   & 45.6\%   & 38.6\%   \\
0.01  & $<$0.001 & $<$0.001 & $<$0.001 & 35.8\%   & 27.0\%   & 10.2\%   \\
0.001 & $<$0.001 & $<$0.001 & $<$0.001 & 0.4\%    & 0.0\%   & 0.0\% \\ \hline
\end{tabular}%
\begin{tablenotes}
\item[1] {\footnotesize Negative cost gap (\%) refers to the percentage of samples with a cost by the proposed algorithm lower than that under the Nash equilibrium.}
\end{tablenotes}
\end{threeparttable}
\end{table}

The distributions of the converged bids under $\sigma_{\gamma}=0.1$ and $a=10$ kW/\$ are illustrated in Fig.~\ref{fig:fidelity-violin}. Although the converged bids may deviate from the Nash equilibrium, their average values equal the Nash equilibrium, validating Theorem \ref{thm-2}.

\section{Conclusion}
\label{sec:conclusion}
This paper proposes a differentially private energy trading mechanism for prosumers in P2P markets, addressing the critical challenge of privacy preservation in distributed energy resource coordination. We formulate the P2P energy trading problem as a generalized Nash game and develop a privacy-preserving Nash equilibrium seeking algorithm that incorporates Laplacian noise to achieve $\epsilon$-differential privacy. Theoretical analysis proves that the proposed algorithm converges in expectation to the Nash equilibrium of the P2P energy trading game. Numerical experiments demonstrate the effectiveness of the proposed algorithm and reveal several interesting findings:
\begin{enumerate}
    \item Increased attack budget may not always lead to higher inference accuracy since the added noise accumulated through iterations.
    \item The added noise could offset the adverse impact of strategic interactions between prosumers, leading to a lower total cost.
\end{enumerate}

Future research directions include exploring adaptive privacy mechanisms, investigating the impact of network constraints on privacy-preserving algorithms, and extending the framework to dynamic market settings with time-coupled constraints.

\appendices

\section{Proof of Proposition~\ref{thm-adversarial}}
\label{appendix:thm-adversarial}
\setcounter{equation}{0}  
\renewcommand{\theequation}{A.\arabic{equation}}

First, the objective function of \eqref{eq:infer} is strictly convex in $\{z_i(k),\forall k=\hat{k}_1,...,\hat{k}_2\}$ and its feasible set is a convex set. Therefore, the optimal $\{z_i(k),\forall k=\hat{k}_1,...,\hat{k}_2\}$ is unique. Furthermore, the objective \eqref{eq:infer-1} is non-negative and setting all variables equal to their actual values results in a feasible solution with a zero objective value. Therefore, at the optimum, we have $z_i(k)=\bar y_i(k),\forall k=\hat{k}_1,...,\hat{k}_2$.

Next, without loss of generality, suppose $i=1$ and $K = \hat{k}_1 - \hat{k}_2 + 1$, we prove that given $z_i(k)=\bar y_i(k),\forall k=\hat{k}_1,...,\hat{k}_2$, $\beta_i$ can be uniquely determined by \eqref{eq:infer-2}-\eqref{eq:infer-4} when $\hat{k}_2-\hat{k}_1 = K - 1 \ge I$ ($I = |\mathcal{I}|)$ and the communication network is fully connected. 

Let $x=[\beta_1, z_2(\hat{k}_1), \cdots, z_I(\hat{k}_1), \cdots, z_2(\hat{k}_2),  \cdots, z_I(\hat{k}_2)]$. From~\eqref{eq:infer-3}, for each iteration $k$, we have
\begin{align}\label{eq:mapping-x}
    z_l(k+1) = q_l z_l(k) + \omega\sum_{j\in N_{\text{com}}^l}z_j(k) + \alpha f_l \hat{\beta}_l, \forall l \in \mathcal{I}
\end{align}
where
\begin{align}
    q_l &= 1 -\omega(I-1)-H_l \nonumber\\
    H_l &= \alpha f_l f_l^\top \nonumber
\end{align}

In the following, we use~\eqref{eq:mapping-x} to construct $\mathbf{A}x = \mathbf{b}$. By verifying matrix $\mathbf{A}$ is of full row rank, we prove $\beta_i$ can be uniquely determined.

Due to the complexity of the matrix, here we use an inductive method: We begin with the constraints for $\hat{k}_1$ and then incrementally add the constraints~\eqref{eq:mapping-x} for the following $k$. Denote the matrix $\mathbf{A}$ as $\mathbf{A}_n$ with constraints for the first $n$ iterations (from $\hat{k}_1$ to $\hat{k}_1 + n - 1$). We verify the matrix $\mathbf{A}_n$ in each step is of full row rank until $n = \hat{k}_2 - 1$, where all constraints are considered and $\mathbf{A}_n = \mathbf{A}$.

When $n = 1$,
\[
\mathbf{A}_1 =
\begin{bmatrix}
\alpha f_1 & \omega \mathbf{1}_{I-1}^\top & \mathbf{0}^\top  & \mathbf{0}^\top\\
\mathbf{0} & \mathbf{M}_q & -\mathbf{I}_{I-1} & \mathbf{0}\\
\end{bmatrix} \in \mathbb{R}^{I \times (K(I - 1) + 1)}
\]
where
\[
\mathbf{M}_q = 
\begin{bmatrix}
q_2 & \omega & \cdots & \omega \\
\omega & q_3 & \cdots & \omega \\
\vdots & \vdots & \ddots & \vdots \\
\omega & \omega & \cdots & q_I
\end{bmatrix} \in \mathbb{R}^{(I - 1) \times (I - 1)}
\]  is a symmetric matrix with diagonal elements \(q_2, \ldots, q_I\) and off-diagonal elements \(\omega\).

Subtracting the first row from the remaining rows, $\mathbf{A}_1$ can be transformed into
\[
\mathbf{A}'_1 = 
\begin{bmatrix}
\alpha f_1 & \omega \mathbf{1}_{I-1}^\top & \mathbf{0}^\top  & \mathbf{0}^\top\\
-\alpha f_1 \mathbf{1} & \mathbf{M}_{q-} & -\mathbf{I}_{I-1} & \mathbf{0}\\
\end{bmatrix}
\]
where
\begin{align}
\mathbf{M}_{q-}=\textbf{diag}(q_2-\omega, q_3 -\omega, \dots, q_I-\omega) \in \mathbb{R}^{(I - 1) \times (I - 1)}.
\end{align}

Then, to eliminate the $\omega \mathbf{1}^\top_{I-1}$ above $\mathbf{M}_{q-}$, we subtract from the first row the sum of $\frac{\omega}{q_i - \omega}$ times the $i$-th row for the remaining rows. Hence, we have

\[
\mathbf{A}''_2 = 
\begin{bNiceArray}{cc:cc}[first-col]
1 &\Delta & \mathbf{0}^\top & \mathbf{v}_q^\top  &  
 \mathbf{0}^\top\\ 
I - 1 & -\alpha f_1 \mathbf{1} & \mathbf{M}_{q-} & -\mathbf{I}_{I-1} & \mathbf{0}\\
\end{bNiceArray}
\]

where
\bsq
\begin{align}
\Delta = \alpha f_1 (1 + \sum_{i = 2}^I \frac{\omega}{q_i-\omega}) \\
\mathbf{v}_q^\top = [\frac{\omega}{q_2-\omega}, \frac{\omega}{q_3-\omega}, \dots, \frac{\omega}{q_I-\omega}]
\end{align}
\esq
Let $\mathbf{B}_1$ and $\mathbf{B}'_1$ denote the left and right sections of $\mathbf{A}''_1$, separated by dashed lines. Since $\mathbf{B}_1$ is of full row rank $I$, $[\mathbf{B}_1, \mathbf{B}'_1]$ is of also full row rank $I$. Moreover, since $\mathbf{B}_1$ is a square matrix, it can be transformed into a diagonal form
$
\mathbf{B}_1 =
\begin{bmatrix}
     \alpha f_1 & \mathbf{0} \\ \mathbf{0} & \mathbf{c}_1
\end{bmatrix}
$, which correspondingly alters $\mathbf{B}'_1$ into $\begin{bmatrix}
    d_2 & \mathbf{0}^\top \\ \mathbf{c}_2 & \mathbf{0}
\end{bmatrix}$. Here, the $\alpha f_1$ in the $\mathbf{B}_1$ because it is one of the eigenvalues.

Next, when $n = 2$, we append $I$ rows (constraints for $k = 2$ in~\eqref{eq:mapping-x}) to $\mathbf{A}_1$, with the similar elementary operations, we have

\NiceMatrixOptions{code-for-first-row = \scriptstyle,code-for-first-col = \scriptstyle }
\[
\mathbf{A}_2 = 
\begin{bNiceArray}{cc:c:cc}[first-col, columns-width=auto]
    1 & \alpha f_1 & \mathbf{0}^\top & \mathbf{d}_2 & \mathbf{0}^\top & \mathbf{0}^\top\\
    I - 1 & \mathbf{0} & \mathbf{c}_1 & \mathbf{c}_2 & \mathbf{0} & \mathbf{0} \\ \hdottedline
    1 & \alpha f_1 & \mathbf{0} & \mathbf{0} & d_2 & \mathbf{0}\\
    I -1 & \mathbf{0} & \mathbf{0} & \mathbf{c}_1 & \mathbf{c}_2 & \mathbf{0} \\
\end{bNiceArray}
\]
We can temporarily omit the ($I+1$)-th row (i.e., the row with $\alpha f_1$). Thus, we have

\[
\Tilde{\mathbf{A}}_2 = 
\begin{bNiceArray}{cc:c:cc}[first-col, columns-width=auto]
    1 & \alpha f_1 & \mathbf{0}^\top & \mathbf{d}_2 & \mathbf{0}^\top & \mathbf{0}^\top\\
    I - 1 & \mathbf{0} & \mathbf{c}_1 & \mathbf{c}_2 & \mathbf{0} & \mathbf{0} \\ \hdottedline
    I -1 & \mathbf{0} & \mathbf{0} & \mathbf{c}_1 & \mathbf{c}_2 & \mathbf{0} \\
\end{bNiceArray}
\]
of size $(1 + 2(I-1)) \times (K(I-1) + 1)$.

Similarly, $\Tilde{\mathbf{A}}_2$ can be represented as $[\mathbf{B}_2, \mathbf{B}'_2]$ and both $\mathbf{B}_2$ and $\Tilde{\mathbf{A}}_2$ are of full rank $\left(1 + 2(I-1)\right)$. Since $\mathbf{B}_2$ is square, it is diagonalizable. Hence, we can recursively use the above operations and the matrix $\mathbf{A}_n$ is of full rank in each step.

By incrementally adding rows for following $k$ with matrix operations, when $n = \hat{k}_2 - 1$, i.e., all constraints for $K - 1$ iterations are considered, the resulting $\Tilde{\mathbf{A}}_n$ is a square matrix of full row rank $1 + K(I-1)$, which is exactly the number of the unknown vector $x$. In addition, the omitted rows satisfy the constraints after checking. This completes the proof. $\hfill \square$

\section{Proof of Theorem~\ref{thm-1}}
\label{appendix:thm-1}
\setcounter{equation}{0}  
\renewcommand{\theequation}{B.\arabic{equation}}

\begin{lemma}[Post-Processing \cite{dwork2014algorithmic}]\label{lemma-1}
Let $\mathcal{M}$ be a randomized
algorithm that is $\varepsilon$-differentially private. Let $\phi~:~\mathbb{R}\to \mathbb{R}'$ be an arbitrary deterministic function. Then $\phi \circ M$ is $\varepsilon$-differentially private.
\end{lemma}

The iterations \eqref{eq:update-privacy} in Algorithm \ref{algo-2} can be written in a compact form as follows:
\begin{align}\label{eq:mapping}
    y(k+1) = Q y(k) + \alpha F (\beta + \gamma), \forall k=0,...,K,
\end{align}
where
\bsq
\begin{align}
    Q= ~ &  \textbf{I}_{I^2} - \omega (\textbf{diag}(|N_\text{com}^1, ..., N_\text{com}^I|)-W) \otimes \textbf{I}_{I} \nonumber \\
    ~ & - \alpha F F^{\top}\label{matrix:Q}\\
     F = ~&  \begin{bmatrix} f_1 & &  \\ & \ddots & \\  & & f_I \end{bmatrix}, \quad \beta = \begin{bmatrix} \beta_1 \\ \vdots \\ \beta_I  \end{bmatrix}, \quad \gamma = \begin{bmatrix} \gamma_1 \\ \vdots \\ \gamma_I \end{bmatrix}.
\end{align}
\esq
where $\otimes$ is the Kronecker product of two matrices. $W \in \mathbb{R}^{I \times I}$ with $W_{ij} = 1$ if $j \in N_\text{com}^i$, otherwise $W_{ij}=0$. Let $\tilde{\mathcal{W}}:= \omega (\textbf{diag}(|N_\text{com}^1|, ..., |N_\text{com}^I|)-W)$. 

First, we prove that the mapping $\tilde{\mathcal{M}}(\beta)=\beta+\gamma$ is $\varepsilon$-differential private for any $\mu$-adjacent $d, d'$, when $\sigma_{\gamma} \ge A\mu/\varepsilon $. Given two $\mu$-adjacent datasets $d, d' \in \mathbb{R}^I$, we have
$\beta, \beta' \in \mathbb{R}^I$ are $(A\mu)$-adjacent. Let $S \subseteq \text{Range}(\tilde{\mathcal{M}}(\beta))$. Let $p_{\beta}(z)$ be the probability density function of $\tilde{\mathcal{M}}(\beta)$ and $p_{\beta'}(z)$ be the probability density function of $\tilde{\mathcal{M}}(\beta')$. We compare the two at some arbitrary point $z \in \mathbb{R}^{I}$ as follows
\begin{align}
    \frac{p_{\beta}(z)}{p_{\beta'}(z)} = ~ &  \prod_{i=1}^I \left(\frac{\text{exp}(-\frac{|z_i-\beta_i|}{\sigma_{\gamma}})}{\text{exp}(-\frac{|z_i-\beta_i'|}{\sigma_{\gamma}})}\right) \nonumber\\
    = ~ & \prod_{i=1}^I \text{exp} \left(\frac{-|z_i-\beta_i|+|z_i-\beta_i'|}{\sigma_{\gamma}}\right) \nonumber\\
    \le ~ & \prod_{i=1}^I \text{exp} \left(\frac{|\beta_i-\beta_i'|}{\sigma_{\gamma}}\right)  \le  \text{exp}(A\mu/\sigma_{\gamma}) \le \text{exp}(\varepsilon)
\end{align}
Therefore,
\begin{align}
    \mathbb{P}[\tilde{\mathcal{M}}(\beta)\subseteq S] \le \text{exp}(\varepsilon) \mathbb{P}[\tilde{\mathcal{M}}(\beta')\subseteq S] 
\end{align}

Following this, let $\phi$ denote the mapping from $\beta+\gamma$ to $y(K)$ given by \eqref{eq:mapping}, according to Lemma \ref{lemma-1}, $\phi \circ \tilde{\mathcal{M}}: \mathbb{R}^I \to \mathbb{R}^I$ from $\beta$ to $y(K)$ is also $\varepsilon$-differential private. This completes the proof. $\hfill \square$

\section{Proof of Theorem~\ref{thm-2}}
\label{appendix:thm-2}
\setcounter{equation}{0}  
\renewcommand{\theequation}{C.\arabic{equation}}

\begin{lemma}\label{lemma-2}
Suppose the spectral radius of matrix $Q$ is $m$, then we have $0<m<1$.
\end{lemma}

\emph{Proof of Lemma \ref{lemma-2}}. First, we define two parameters
\begin{align}
    m_1 = \frac{\overline{\lambda}+4 \alpha \overline{f}^2 + \sqrt{\overline{\lambda}^2 + 4 \alpha ^2 \overline{f}^4}}{2}-1 \\
     m_2 = \frac{-\underline{\lambda}-2\alpha \underline{f}^2 + \sqrt{\underline{\lambda}^2 + 4 \alpha^2 \overline{f}^4}}{2}+1
\end{align}

It is easy to prove that when assumption A1 holds, $m_1<1$ and $m_2<1$.

Then, we prove the spectral radius of matrix $Q$ is less than 1 by proving that $Q+m_1 \textbf{I}_{I^2}>0$ and $m_2 \textbf{I}_{I^2} -Q>0$ as follows:

Let $\tilde{\mathcal{W}}:= \omega (\textbf{diag}(|N_\text{com}^1|, ..., |N_\text{com}^I|)-W)$. Denote the eigenvalues of $\tilde{\mathcal{W}}$ as $\lambda_1, \lambda_2, ..., \lambda_I$. Obviously, $0$ is one of the eigenvalues and without loss of generality, we assume $\lambda_1=0$. Denote the corresponding orthogonal unit eigenvectors by $u_1, u_2, ..., u_I$, respectively. Let $U_1=u_1 \otimes \textbf{I}_I$ and $U_2 = [u_2, ..., u_I] \otimes \textbf{I}_I$.

First we check
\begin{align}
 [U_1, U_2]^{\top} \left( Q+ m_1 \textbf{I}_{I^2} \right) [U_1, U_2] = \begin{bmatrix}
     \Pi_1 & \Pi_2\\
     \Pi_3 & \Pi_4
  \end{bmatrix}
\end{align}
where
\bsq
\begin{align}
    \Pi_1 =~ & (1+m_1) \textbf{I}_{I^2} - \alpha U_1^{\top} F F^{\top} U_1 \\
    \Pi_2 = ~ & -\alpha U_1^{\top} F F^{\top} U_2 \\
    \Pi_3 = ~ & -\alpha U_2^{\top} F F^{\top} U_1 \\
    \Pi_4 = ~ & (1+m_1) \textbf{I}_{I^2} - \textbf{diag}(\lambda_2, ...,\lambda_I)-\alpha U_2^{\top} F F^{\top} U_2
\end{align}
\esq

Since
\begin{align}
   & \Pi_1 \ge 1+m_1 - \alpha \overline{f}^2 >0, \label{eq:eig-1}\\
   & \Pi_4 - \Pi_3 \Pi_1^{-1} \Pi_2 \ge 1+m_1 - \overline{\lambda} - \alpha \overline{f}^2 - \frac{\alpha^2 \overline{f}^4}{1+m_1-\alpha \overline{f}^2}=0, \label{eq:eig-2}
\end{align}
matrix $Q+m_1 \textbf{I}_{I^2}>0$. \eqref{eq:eig-1} and \eqref{eq:eig-2} are because of the definition of $m_1$.

Similarly, we check
\begin{align}
    [u_1, U_2]^{\top} (m_2 \textbf{I}_{I^2}-Q) [U_1, U_2] = \begin{bmatrix}
        \Pi_1' & \Pi_2'\\ \Pi_3' & \Pi_4'
    \end{bmatrix}
\end{align}
where
\bsq
\begin{align}
    \Pi_1' = ~ & (m_2 - 1) \textbf{I}_{I^2}+ \alpha U_1^{\top} F F^{\top} U_1 \\
    \Pi_2' = ~ & \alpha U_1^{\top} F F^{\top} U_2\\
    \Pi_3' = ~ & \alpha U_2^{\top} F F^{\top} U_1\\
    \Pi_4' = ~ & (m_2-1) \textbf{I}_{I^2} + \textbf{diag}(\lambda_2, ...,\lambda_I)+\alpha U_2^{\top} F F^{\top} U_2
\end{align}
\esq

Since
\begin{align}
    & \Pi_1' \ge m_2-1 + \alpha \underline{f}^2 >0, \label{eq:eig-3} \\
    & \Pi_4' - \Pi_3' \Pi_1'^{-1} \Pi_2' \ge m_2 -1 +\underline{\lambda}+ \alpha \underline{f}^2 - \frac{\alpha^2 \overline{f}^4}{m_2-1 + \alpha \underline{f}^2} =0, \label{eq:eig-4}
\end{align}
matrix $m_2 \textbf{I}_{I^2}-Q >0$. \eqref{eq:eig-3} and \eqref{eq:eig-4} are because of the definition of $m_2$.

Based on the above analysis, the maximum eigenvalue of the matrix $Q$, $m<\max\{|m_1|, |m_2|\}<1$, or equivalently, the spectral radius of matrix $Q$ is strictly less than 1.

$\hfill \square$

With Lemma \ref{lemma-2}, we continue to prove Theorem \ref{thm-2}. Recall that the iterations in Algorithm \ref{algo-2} are given by \eqref{eq:mapping}. Denote the error between $y(k)$ and the Nash equilibrium $b^*$ by $\delta(k)$, i.e.,
\begin{align}
    \delta(k) = y(k) - \textbf{1}_I \otimes b^*
\end{align}
Moreover, based on the convergence of Algorithm \ref{algo-1}, we have
\begin{align}
    \textbf{1}_I \otimes b^* = Q ( \textbf{1}_I \otimes b^*) + \alpha F \beta
\end{align}
Hence, the change of $\delta(k)$ is given by 
\begin{align}\label{eq:delta}
    \delta(k+1) = Q \delta (k) + \alpha F \gamma
\end{align}
Summing \eqref{eq:delta} up from $0$ to $k-1$ yields
\begin{align}
    \delta(k) = Q^k \delta(0) + \alpha \sum_{j=0}^{k-1} Q^j F \gamma
\end{align}

Since the eigenvalues of the matrix $Q$ is strictly less than 1 and we have $\mathbb{E}[\gamma]=0$, so
\begin{align}
    \lim_{k \to \infty} \mathbb{E}[\delta(k)] = \lim_{k \to \infty} \mathbb{E}[Q^{k} \delta(0)] =0.
\end{align}
Thus, $\lim_{k \to \infty} \mathbb{E}[y_i(k)] = b^*,\forall i \in \mathcal{I}$.

Furthermore, we evaluate the value of $\mathbb{E}||\delta(k)||^2$ as follows.
\begin{align}
    \mathbb{E}||\delta(k+1)||^2 = ~ & \mathbb{E}||Q\delta(k)+\alpha F\gamma||^2 \nonumber\\
    \le ~ & \mathbb{E}||Q\delta(k)||^2+ \alpha^2 \mathbb{E}|| F\gamma||^2 \nonumber\\
    \le ~ & ||Q||^2 \mathbb{E}||\delta(k)||^2 + 2\alpha^2 \text{Tr}(F^{\top}F) \sigma_{\gamma}^2 
\end{align}
According to Lemma \ref{lemma-2}, we have $||Q||=m <1$. Hence,
\begin{align}
    \mathbb{E}||\delta(k)||^2 \le ~ &  (m^k)^2 \mathbb{E}||\delta(0)|| + \frac{2\alpha^2 \text{Tr}(F^{\top}F) \sigma_{\gamma}^2}{1-m^2}
\end{align}
and
\begin{align}
    \lim_{k \to \infty} \mathbb{E}||y(k)-\textbf{1}_I \otimes b^*||^2 = ~ & \lim_{k \to \infty} \mathbb{E}||\delta(k)||^2 \nonumber\\
   \le ~ &  \frac{2\alpha^2 I  \sigma_{\gamma}^2 \max\{||f_i||^2,\forall i\}}{1-m^2}
\end{align}

This completes the proof. $\hfill \square$

\bibliographystyle{IEEEtran}
\bibliography{mybib}


\end{document}